\documentclass[%
reprint,
superscriptaddress,
noeprint,
amsmath,amssymb,
aps,pra,
]{revtex4-2}

\usepackage{graphicx}
\usepackage{dcolumn}
\usepackage{bm}
\usepackage{xcolor} 
\usepackage{hyperref}
\usepackage{siunitx}
\DeclareSIUnit\bohr{\text {\ensuremath {a}}_{0}}
\DeclareSIUnit\gauss{\text{G}}

\usepackage{xspace}
\usepackage{braket}
\usepackage{placeins}

\usepackage{tikz}
\newcommand*\circled[1]{\tikz[baseline=(char.base)]{
		\node[shape=circle,draw,inner sep=1pt] (char) {#1};}}

\newcommand{\Cs}{$^{133}$Cs\xspace}
\newcommand{\Li}{$^6$Li\xspace}

\begin{document}
	
	\preprint{APS/123-QED}
	
	\title{The heavy Fermi polaron I: the Lithium-Cesium experiment}%
	
	\author{Michael Rautenberg}
	\thanks{These authors contributed equally to this work.}
	\author{Tobias Krom}%
	\thanks{These authors contributed equally to this work.}
	\affiliation{%
		Physikalisches Institut, Universität Heidelberg, 69120 Heidelberg, Germany
	}%
	
	\author{Eleonora Lippi}
	\thanks{present address: Department of Physics and Research Center OPTIMAS, RPTU University Kaiserslautern-Landau, 67663 Kaiserslautern, Germany}
	\affiliation{%
		Physikalisches Institut, Universität Heidelberg, 69120 Heidelberg, Germany
	}%
	
	\author{Lauriane Chomaz}
	\email{chomaz@uni-heidelberg.de}
	\author{Matthias Weidemüller}%
	\email{weidemueller@uni-heidelberg.de}
	\affiliation{%
		Physikalisches Institut, Universität Heidelberg, 69120 Heidelberg, Germany
	}%
	
	\date{\today}
	
	\begin{abstract}
		We present details of an experimental platform for studying Fermi polarons in a quantum-gas mixture. The system consists of about a thousand bosonic $^{133}$Cs impurities immersed in a deeply degenerate Fermi gas ($T/T_F \sim 0.2$) of approximately \num{2e5} $^6$Li atoms in a single hyperfine state, with interspecies interactions tunable via a Feshbach resonance. Using optical Raman spectroscopy without relative momentum transfer, we perform injection spectroscopy and thereby create the Fermi polaron. Owing to the large mass imbalance between the two species, the setup provides access to previously unexplored regimes of Fermi polarons.
	\end{abstract}

	\maketitle

	\section{Introduction}

	Since polarons were introduced in the context of solid-state physics \cite{Landau1948_Effective, Frohlich1954_Electrons}, the concept has since been extended beyond its original setting. Today, the term polaron broadly refers to a quasiparticle consisting of an impurity dressed by excitations of a surrounding quantum bath. This generalized framework has led to extensive studies of polaron physics, particularly in ultracold atomic gases \cite{Massignan2014_Polarons, Scazza2022_Repulsive, Grusdt2025_Impurities, Parish2025_Fermi}.
	Specifically to describe the Fermi polaron, where the quantum bath is an initially non-interacting Fermi sea, multiple theoretical models have been developed.
	For the equal mass case, \textit{i.e.} the impurity mass $M$ equal to the mass of the surrounding Fermions $m$, variational methods work remarkably well \cite{Chevy2006_Universal, Parish2013_Highlya, Liu2020_RadioFrequency}.
	In the other extreme case of an infinitely heavy impurity $M/m\to\infty$, the system can be solved exactly \cite{Knap2012_TimeDependent,Schmidt2018_Universal, Wang2023_Functional}
	and exhibits the so-called Anderson Orthogonality Catastrophe in the thermodynamic limit \cite{Anderson1967_Infrared}.
	
	While various physical systems feature polaronic quasiparticles, cold atom experiments offer pristine conditions to study polarons experimentally, due to the high degree of control over internal states and interactions of impurity and bath particles \cite{Baroni2024_Quantum, Massignan2026_Polarons}. Since the first time when degenerate Fermi gases could be reliably prepared in cold atom experiments, impurity physics was studied using two distinct internal states of the fermions as impurity and bath particles, resulting in the observation of an equal-mass Fermi polaron \cite{Schirotzek2009_Observation, Scazza2017_Repulsive, Ness2020_Observation}.
	So far, the experimental studies have been extended towards heavier impurities immersed in a degenerate Fermi gas by the Grimm group in Innsbruck only, using $^{40}\mathrm{K}$-\Li and $^{41}\mathrm{K}$-\Li mixtures with a mass ratio of $M/m \approx7$ \cite{Kohstall2012_Metastability, Cetina2015_Decoherence, Fritsche2021_Stability, Baroni2024_Mediated}.

	To extend the study to even heavier impurities, we realize a \Li-\Cs mixture with a mass imbalance of $M/m \approx 22$. In this regime, theories of Fermi polarons with a very heavy, yet finite mass impurity, recently revealed connections of the Anderson orthogonality catastrophe for a static impurity on the one hand to the quasiparticle picture of mobile impurities on the other hand \cite{Chen2025_MassGap}.
	Along with tunable interspecies interactions provided by magnetic Feshbach resonances, the \Li-\Cs system constitutes a unique platform to study this regime. Previously, other kinds of few- and many-body effects have been explored with this ultracold mixture such as Efimov physics \cite{Pires2014_Observation, Tung2014_Geometric, Johansen2017_Testing, Ulmanis2016_Heteronuclear, Tran2021_Fermions} and mediated interactions \cite{Desalvo2019_Observation, Cai2026_Fermion}.
	
	In this article we describe in detail
	the experimental techniques needed to investigate the heavy Li-Cs Fermi polaron spectroscopically. The basic parts of the experiment, including the fundamental laser cooling routines, can be found in \cite{Lippi2024_Experimental}.
	In Sec.~\ref{sec:sequence} the preparation of the degenerate Fermi gas of \Li with embedded \Cs impurities is outlined. Sec.~\ref{sec:Raman} contains details on the Raman spectroscopy used to record the polaron spectra, which are then presented in Sec.~\ref{sec:spectra}. 
	
	\section{Cesium impurities in a Lithium Fermi sea}   \label{sec:sequence}
	
\begin{figure}
		\centering
		\includegraphics[width=1\linewidth]{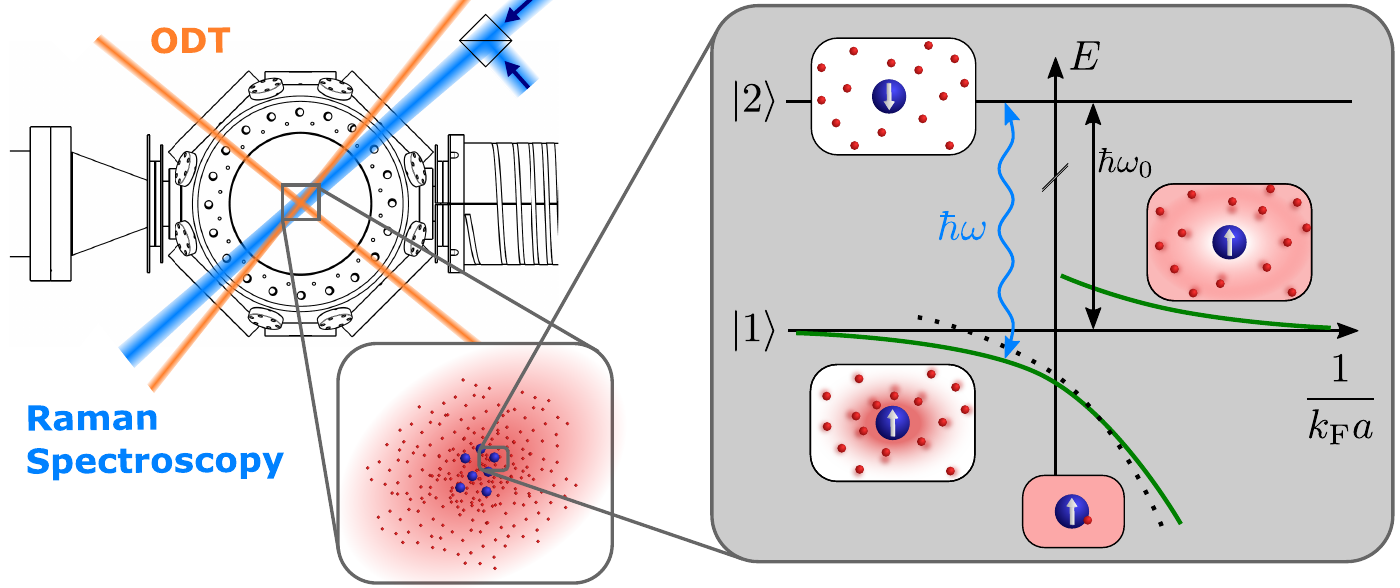}
		\caption{Left: Major components of the experimental setup (final optical dipole trap, ODT, and two co-propagating Raman spectroscopy beams) and schematic representation of the resulting sample (bottom left) of a few Cs impurities (blue) in a dense quantum degenerate Fermi sea of Li (red). Right box: Level scheme relevant for the spectroscopy using two internal impurity states. The impurity Cs$\ket{1}$ state features tunable interactions with the surrounding bath atoms and forms a polaron, whereas the effectively non-interacting Cs$\ket{2}$ state serves as a reference state. Depending on the impurity-bath interaction parameter $1/(k_\mathrm{F}a)$, the system can either be in the attractive (negative energies) or repulsive (positive energies) polaron state (solid green lines). In addition, a dressed dimer (dotted black line) state can be formed. The energies of these states can be measured spectroscopically as a deviation of the two-photon transition frequency $\omega$ from the bare resonance frequency $\omega_0$.}
		\label{fig:Fig1}
	\end{figure}

	In our experiment, we aim for the creation of a large Fermi sea of \Li atoms in a single spin state mixed with a much smaller cloud of \Cs impurities, \textit{cf.} Fig.~\ref{fig:Fig1}. 
	The combination of two different atomic species, such as Li and Cs, imposes certain constraints on the preparation of the desired sample. Due to their different polarizabilities as well as the minimal achievable temperatures after laser cooling, we cannot simultaneously load Li and Cs into a common optical dipole trap from a double-species magneto-optical trap (MOT). Instead, we prepare the sample sequentially. First, Cs is loaded into its dipole trap displaced from where afterwards the Li atoms are loaded into a MOT and then into their respective dipole trap. Only after considerable forced evaporation, both clouds can be combined in a single optical trap. 
	
\begin{figure*}
		\centering
		\includegraphics[width=0.75\linewidth]{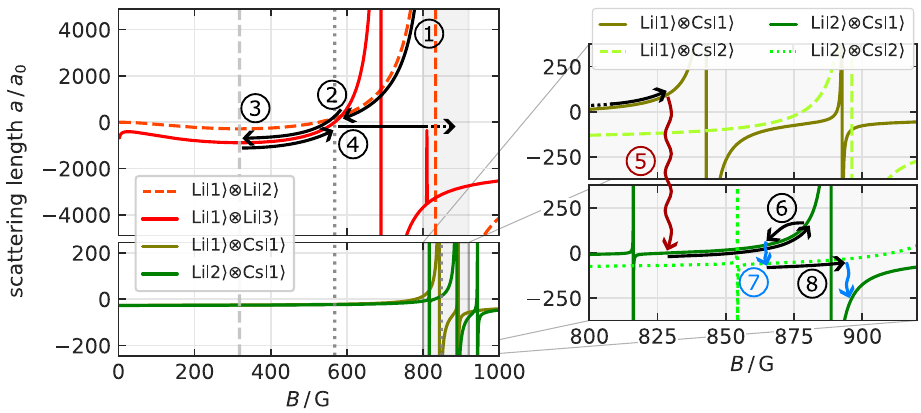}
		\caption{Magnetic field dependence of intra- (upper panel) and interspecies (lower panel) interactions relevant for the sample preparation parametrized by their respective $s$-wave scattering lengths $a$. The gray dashed line marks the magnetic field where the Li$\ket{1}$-Li$\ket{3}$ \cite{Note1} mixture is evaporated, whereas the gray dotted line marks the field where Li$\ket{1}$-Li$\ket{3}$ interactions vanish. A detailed view around the Li$\ket{2}$-Cs$\ket{1}$ Feshbach resonance close to \SI{888.6}{\gauss} used for the polaron spectroscopy is presented  on the right-hand side. The upper panel shows the Li$\ket{1}$-Cs scattering lengths, whereas the lower panel contains those of Li$\ket{2}$-Cs. Black arrows indicate magnetic field ramps during sample preparation whereas wavy lines symbolize Li radio-frequency (red) or Cs Raman transitions (blue) between different hyperfine states. Circled numbers correspond to preparation steps referenced in the main text.}
		\label{fig:FB_resonances}
	\end{figure*}

	Another major point to consider is the structure of intra- and interspecies magnetic Feshbach resonances (FR), which govern the scattering properties of the different Li and Cs states \textit{cf.} Fig.~\ref{fig:FB_resonances}. The Li-Cs FR that we use for the polaron experiments is located at \SI{888.6}{\gauss}. Efficient evaporative cooling of Li requires a mixture of two different hyperfine states with strong scattering, \textit{e.g.} provided by the broad Li$\ket{1}$-Li$\ket{2}$ \footnote{Throughout this work, the spin states of Li and Cs are labeled according to their energies at finite magnetic fields, starting with the ground state $\ket{1}$} FR close to \SI{832}{\gauss} \cite{Ketterle2008_Making, Hulet2020_Methods}.
	
	As we want to create a deeply degenerate Fermi gas of Li in a single hyperfine state, the removal of the atoms in the other populated state is required. We achieve this by heating the atoms out of the trap using a resonant light pulse. Interactions between the two Li states involved during this light pulse leads to heating on the remaining atoms. Such collision processes can be avoided when removing the atoms at vanishing interactions. Unfortunately, evaporating close to the \SI{832}{\gauss} Feshbach resonance and subsequently ramping the magnetic field to the zero crossing near \SI{527}{\gauss} would inevitably populate the Li-Li Feshbach dimer state, leading to losses through three-body recombination in the non-universal regime near the interaction zero \cite{Lompe2011_phd}. This issue cannot be circumvented with a fast magnetic field ramp, due to the large width of the Li FR. 
	Thus, we need to evaporate the sample at a field from which we can approach the zero of the scattering length via the attractive side where no dimer state exists.
	
	To comply with these restrictions, we use a Li$\ket{1}$-Li$\ket{3}$ mixture to evaporate near \SI{320}{\gauss}  after a short initial evaporation in the Li$\ket{1}$-Li$\ket{2}$ mixture near \SI{832}{\gauss}, as shown as \circled{1} in Fig.~\ref{fig:FB_resonances}. The Li$\ket{1}$-Li$\ket{3}$ scattering at \SI{320}{\gauss} is sufficient for evaporative cooling and transferring Li into the final optical dipole trap (ODT) \circled{3} involving an adiabatic state transfer at \circled{2}. 
	From there, we ramp the magnetic offset field to $\approx \SI{568}{\gauss}$ where Li$\ket{1}$-Li$\ket{3}$ is non-interacting, allowing Li$\ket{3}$ to be removed with a resonant light pulse \circled{4}. The remainder of this section will provide a detailed description of this sample preparation procedure.
	
	\subsection{Preparation}  \label{subsec:sample_preparation}

	Our sequence starts with loading around \num{1e7} \Cs atoms in a MOT, followed by two stages of laser cooling (sub-Doppler \cite{Drewsen1994_Investigation} and degenerate Raman-sideband cooling \cite{Vuletic1998_Degenerate, Kerman2000_Optical, Treutlein2001_Highbrightnessa}) resulting in a sample of \num{6e6} Cs atoms in their ground state $\ket{F=3, m_F=3} \equiv \mathrm{Cs}\ket{1}$. Details pertaining to these early stages of the Cs preparation can be found in \cite{Lippi2024_Experimental}.
	The Cs atoms are then loaded into a large-volume dipole trap, the so-called ``reservoir trap" (RT), at an offset field of \SI{42}{\gauss}. 
	The RT is implemented as a crossed optical trap with a \SI{90}{\degree} crossing angle with a wavelength of \SI{1064}{\nano\meter} \footnote{Innolight Mephisto Mopa \SI{55}{\watt}} and beam waists of around \SI{350}{\micro\meter},
	where a single beam is recycled after the vacuum chamber to generate the second arm of the trap. Trap loading is facilitated by a magnetic gradient field %
	that creates a potential canceling the effect of gravity for the Cs$\ket{1}$ state, leading to a deeper trap along the direction of gravity, thus guaranteeing a better overlap in momentum space with the laser-cooled Cs cloud. This magnetic levitation is subsequently ramped down while the laser power of the trap is increased in order to minimize Cs losses during the following Li preparation. Note that all the steps above including the loading of the RT happen displaced from the center of the vacuum chamber where Li will be prepared next, so that Cs can be kept apart from Li during its initial loading and evaporation steps. 
	
	Next, in the center of the vacuum chamber, \num{7.5e7} \Li atoms are loaded in a MOT. The early preparation steps of Li are detailed in Ref.~\cite{Lippi2024_Experimental}. Following a MOT compression phase, the atoms are further cooled using a gray molasses scheme on the $D_1$ line \cite{Grier2013_Lambda, Burchianti2014_Efficient} before they are loaded into another dipole trap (DT). The DT is a cigar-shaped trap designed to provide the strong potential needed to trap Li at \SI{70}{\micro \kelvin} after laser cooling. It consists of two beams of around \SI{70}{\micro\meter} beam waist crossing at \SI{7}{\degree}.
	The beams are derived from a \SI{200}{W} fiber laser operating at \SI{1070}{\nano\meter} \footnote{IPG YLR-200-LP-WC}. To increase the loading efficiency, the beam pointing is modulated during loading (using an acousto-optic modulator), leading to a time-averaged potential with a larger capture volume \cite{Ahmadi2005_Geometrical}. After ramping down this modulation, we end up with a roughly equal mixture of \num{1e6} atoms in each of the lowest two hyperfine states, Li$\ket{1}$ and Li$\ket{2}$.
	
\begin{figure}
		\centering
		\includegraphics[width=1\linewidth]{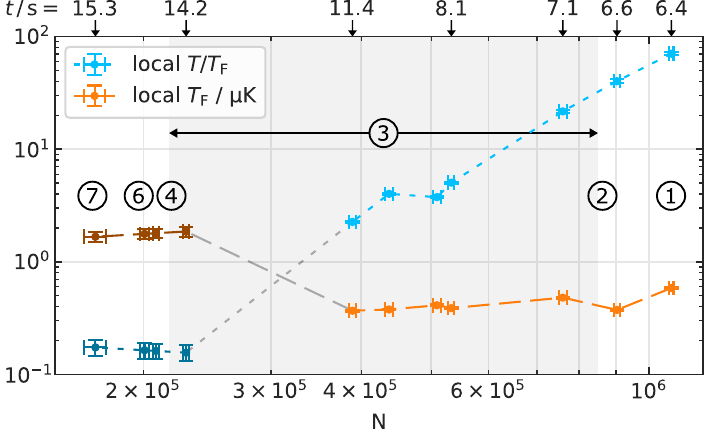}  
		\caption{Local Fermi temperature $T_\mathrm{F}$ and ratio of temperature to $T_\mathrm{F}$ as a function of the number of Li atoms (per spin state, where applicable) over the experimental sequence. Note that time $t$ runs from right to left as indicated by the arrows at the top.
			The circled numbers correspond to the preparation steps in Fig.~\ref{fig:FB_resonances}, and the shaded region marks the evaporation at \SI{320}{\gauss}. 
			The darker (lighter) colored points refer to Li in the final trap (DT).}
		\label{fig:TTF_vs_N}
	\end{figure}

	Fast thermalization close to the broad Li$\ket{1}$ - Li$\ket{2}$ Feshbach resonance around \SI{832}{\gauss} \circled{1} facilitates efficient loading. After the DT modulation is ramped down, Li is further evaporatively cooled by lowering the laser power of the DT, \textit{cf.} Fig.~\ref{fig:TTF_vs_N}. At a temperature of \SI{10}{\micro\kelvin}, at which the number of thermally created Feshbach molecules remains small \cite{Chin2004_Thermal}, an adiabatic radio-frequency (rf) sweep \cite{Ketterle2008_Making} is used to transfer Li$\ket{2}$ to Li$\ket{3}$
	\circled{2}~\cite{Hulet2020_Methods, Lompe2011_phd}.
	This transfer allows the use of the comparably strong Li$\ket{1}$-Li$\ket{3}$ interactions at \SI{320}{\gauss} \circled{3} with a scattering length of \SI{-900}{\bohr}, (with the Bohr radius $\SI{}{\bohr}$). 
	
	Before the removal of the Li$\ket{3}$ atoms, Li evaporation is continued until a temperature of \SI{1.5}{\micro K} at which point the Cs atoms are combined with the Li sample by moving the crossing point of the RT beams to the position of the trapped Li cloud.
	The motion is mainly carried out in the horizontal plane using a piezoelectric mirror mount \footnote{Thorlabs POLARIS-K1S2P}, where the long RT beam path enables displacements of up to \SI{950}{\micro\meter}.
	To prevent unwanted intersections between beams belonging to different dipole traps during the move, an additional vertical shift of \SI{190}{\micro\meter} is applied using two additional actuators.
	
	The trap depths during the transfer are optimized to minimize Li losses and heating due to sudden trap potential deformations. Owing to the approximately fourfold larger trap depth experienced by Cs in the \SI{1070}{\nano\meter} DT, this optimization necessarily compromises the Cs transfer efficiency. Consequently, only \num{3e4} Cs atoms remain in the DT after the transfer.
	Even though both species are now trapped in the same optical dipole trap, they do not thermalize with each other due to the weak Li–Cs interactions $|a|\lesssim\SI{30}{\bohr}$. 
	After a final evaporation ramp of the DT, leaving Li at around \SI{800}{nK}, 
	both species are loaded into the final ODT, which consists of two beams with beam waists of around \SI{45}{\micro\meter} and a crossing angle of \SI{90}{\degree} operating at a wavelength of \SI{1064}{\nano\meter}.
	
	Loading Li from the DT into the final ODT immediately 
	decreases $T/T_\mathrm{F}$ (temperature $T$ over local Fermi temperature $T_\mathrm{F}$, \textit{cf.} Eq.~\ref{eq:local_EF}), similar to an adiabatic phase-space compression \cite{Serwane2011_Deterministic, Viverit2001_Adiabatic} 
	from $T/T_\mathrm{F}\gtrsim 1$ to $T/T_\mathrm{F}\approx 0.2$, essentially avoiding the inefficient evaporation of a Fermi gas in the degenerate regime, see Fig.~\ref{fig:TTF_vs_N}. 
	
	Once the DT is fully turned off, we ramp the offset field to \SI{568}{\gauss} where the Li$\ket{1}$ - Li$\ket{3}$ interactions vanish so that we can remove the atoms in the Li$\ket{3}$ state by shining in a resonant light pulse \circled{4}. 
	After transferring the Li atoms from Li$\ket{1}$ to Li$\ket{2}$ using another adiabatic radio-frequency sweep \circled{5}, we use moderate interactions with $a \approx \SI{170}{\bohr}$ in the wing of the Li$\ket{2}$-Cs$\ket{1}$ \SI{888.6}{\gauss}-resonance to sympathetically cool Cs \circled{6}.  This thermalization process is also used to accurately determine the Li temperature, see Sec.~\ref{subsec:thermometry}. We go again to slightly weaker Li-Cs interactions at \SI{865}{\gauss} to transfer Cs$\ket{1}$ to Cs$\ket{2}$  \circled{7}, using either an adiabatic sweep or a $\pi$ pulse with the Raman spectroscopy setup (see below in Sec.~\ref{sec:Raman}). 
	
	We then proceed to the target field \circled{8} near the Li$\ket{2}$-Cs$\ket{1}$ Feshbach resonance at $B_0 = \SI{888.577(10)(10)}{\gauss}$ \cite{Johansen2017_Testing, Ulmanis2015_Universality}.
	This broad Feshbach resonance is well suited for impurity investigations owing to its large width, $\Delta B = \SI{-58.1}{\gauss}$~\cite{Johansen2017_Feshbach}, 
	which ensures excellent tunability of the interaction strength (see Sec.~\ref{subsec:magnetic_field_stability}), and its small effective range parameter, $R^*=\SI{65.2}{\bohr}$ \cite{Johansen2017_Testing}, 
	which allows to neglect effective-range corrections. Additionally, the magnetic-field range close to this resonance offers an almost non-interacting reference state Cs$\ket{2}$ with approximately constant $a_\mathrm{Li\ket{2}-Cs\ket{2}} \approx \SI{-41}{\bohr}$, which is essential for polaron spectroscopy. 
	
	After a total preparation time of $\gtrsim \SI{20}{\second}$, our final sample consists of \num{2e5} lithium atoms in state Li$\ket{2}$, \num{1000} cesium atoms in state Cs$\ket{1}$, both at a temperature of \SI{300}{nK}. Details on the sample characterization can be found in  later Sec.~\ref{subsec:sample_characterization}. We ultimately find that the average Li density sampled by the Cs impurities is \SI{4.0(7)e12}{\per\cubic\centi\meter} yielding a local Fermi energy $E_\mathrm{F} = h \times \SI{32(4)}{\kilo \hertz}  = k_\mathrm{B} \times \SI{1.53(17)}{\micro \kelvin}$ corresponding to a Li Fermi sea with a local $T/T_\mathrm{F} = \num{0.20(3)}$. The inverse Li Fermi wavevector $1/k_\mathrm{F} = \SI{3060+-170}{\bohr}$ is used in conjunction with the Li$\ket{2}$-Cs$\ket{1}$ scattering length $a$ to define the dimensionless interaction parameter $1/(k_\mathrm{F}a)$. 
	
	\subsection{Characterization} \label{subsec:sample_characterization}
	Since all properties of the Fermi polaron depend on the parameters of the Fermi sea representing the quantum bath, it is essential to precisely know both the local Fermi energy $E_\mathrm{F} = k_\mathrm{B} T_\mathrm{F}$ sampled by the impurity cloud as well as the degree of degeneracy $T/T_\mathrm{F}$. The local Fermi energy 
	\begin{equation} \label{eq:local_EF}
		E_\mathrm{F} = \frac{\hbar^2 \left(6\pi^2 n(\mathbf{r})\right)^{2/3}}{2m_\mathrm{Li}} 
	\end{equation} 
	is a function of solely the fermion density $n(\mathbf{r})$ which in turn can be determined from the total number of Li atoms, the trapping potential, as well as the temperature of the fermions \cite{Ketterle2008_Making}. The remainder of this section thus describes the necessary steps to measure the Li temperature and to determine the parameters of the trap potential including both optical and magnetic potentials.
	To quantify the uncertainty of the impurity-bath interactions parametrized by $1/(k_\mathrm{F}a)$ where the Li$\ket{2}$-Cs$\ket{1}$ $s$-wave scattering length $a$ can be tuned via the magnetic offset field, we will additionally comment on the magnetic field stability in Sec.~\ref{subsec:magnetic_field_stability}.
	
	\subsubsection{Magnetic field topology} \label{sec:mag_topology}
	Due to geometrical constraints in coil placement around the vacuum chamber, the coils providing the magnetic offset field (``Feshbach coils") are not exactly arranged in a Helmholtz configuration, which at our usual working fields around \SI{900}{\gauss} yields a saddle potential $ \vert B \vert \approx B_0 \times (1+\alpha (z^2-(x^2+y^2)/2)$ (confining in-plane and anti-confining along gravity) with $\alpha= 0.0242(9)_\mathrm{stat} (14)_\mathrm{sys}\;\mathrm{cm}^{-2}$ 
	and with significant curvature of the magnetic fields (see also Ref.~\cite{Lippi2024_Experimental}).
	
	Care has been taken to align the center of our final ODT with the center of the Feshbach coils. To determine the Feshbach coil center in-plane, we observe the center of mass motion of a Li cloud trapped in only one beam of our final ODT. To determine the magnetic-field minimum along the direction of gravity, we measure the absolute magnetic field using radio-frequency spectroscopy on Li atoms trapped in the final ODT after moving it to different heights.
	
	\subsubsection{Optical potential}
	
\begin{figure}
		\centering
		\includegraphics[width=1\linewidth]{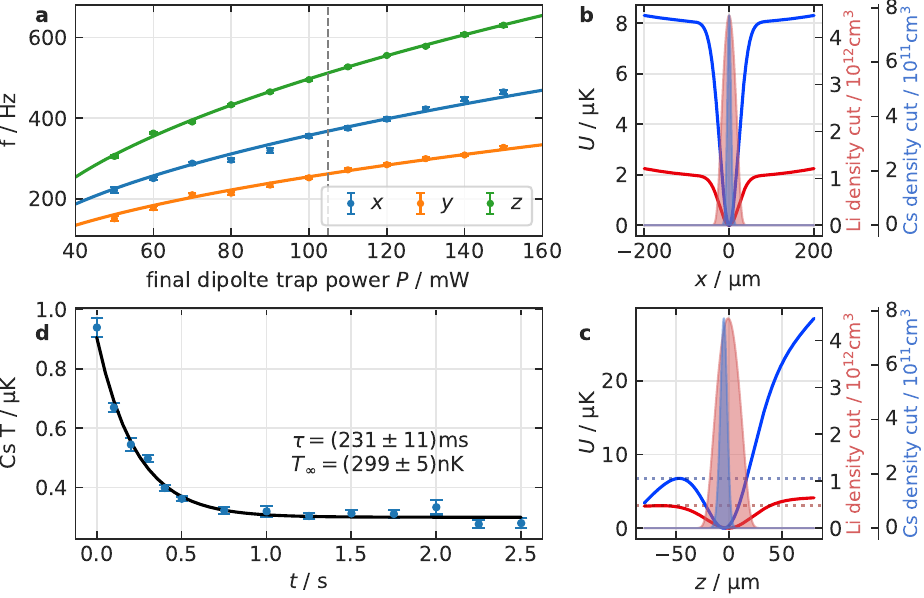}
		\caption{\textbf{a}: Li trap frequencies in the final ODT. Each data point is the frequency resulting from fitting a damped sinusoidal oscillation to the time evolution of the size of a non-interacting cloud of Li atoms following a sudden change in trap power. Error bars represent the standard error of the fit. Solid lines are predictions of the model including optical, magnetic, and gravitational potentials. The colors represent measurements along three spatial directions. The vertical dashed line marks the trap power used in the polaron experiments. 
			\textbf{b} and \textbf{c}: trap potentials for Li (Cs) as red (blue) solid lines together with cuts of the density distributions through the trap minimum (shaded in the same color) in-plane (\textbf{b}) and along gravity (\textbf{c}).
			\textbf{d}: Cs thermalization to the Li temperature at \SI{880}{\gauss}. The Cs temperature is shown as a function of the time $t$ at the target magnetic field together with a fit of the form  $T_{Cs}(t) = A \times \exp(-t/\tau) + T_\infty$. 
		}
		\label{fig:final_sample}
	\end{figure}

	To characterize the potential in which the atoms are trapped, we measure periodic oscillations of the sizes of a cloud of non-interacting Li atoms after a sudden quench in the laser intensity of the trapping beams. Repeating this procedure at different laser intensities and comparing the results to calculated trap frequencies from the trap potential model including magnetic (see Sec.~\ref{sec:mag_topology}), optical, and gravitational potentials allows us to fix the parameters of the optical potential in this model.  
	An optical dipole trap
	formed by elliptical beams with \num{0.88} aspect ratio and horizontal beam waists of \SI{45}{\micro \meter} and \SI{59}{\micro \meter}, respectively, reproduce the trap frequencies over a wide range of laser powers of \qtyrange[range-units = single]{50}{150}{\milli W}, see Fig.~\ref{fig:final_sample}a. For the polaron experiments, we use a power of \SI{105}{\milli\watt} per beam, resulting in the potentials shown in Fig.~\ref{fig:final_sample}b and c, with a mean trap frequency for Li of $2\pi\times$\SI{367(3)}{\hertz}.

	\subsubsection{Thermometry on Lithium Fermi sea} \label{subsec:thermometry}
	
	Since the accurate determination of the temperature of a deeply quantum degenerate Fermi gas from fitting cloud shapes measured via absorption imaging is prone to large systematic errors as described in Appendix~\ref{appendix:Li_thermometry}, we rely on using the Cs impurities as a thermometer \cite{Lous2017_Thermometry, Bouton2020_SingleAtom}.
	
	After a thermalization time of $\gtrsim\SI{500}{\milli\second}$ at moderate interactions (\textit{c.f.} Fig.~\ref{fig:final_sample}d at $a_\mathrm{Li\ket{2}-Cs\ket{1}} \approx \SI{170}{\bohr}$), Cs has reached an equilibrium temperature, which corresponds to the temperature of the surrounding Li bath. The Cs equilibrium temperature is reached much faster than the timescale of atom loss ($>\SI{1}{\second}$), which is dominated by Li-Cs-Cs three-body losses.
	We use a time-of-flight expansion series of Cs to accurately determine the Li temperature. 
	This temperature is consistent with the trap depth $U \approx k_\mathrm{B} \times \SI{3.1}{\micro\kelvin}$ 
	extracted from the model of Sec.~\ref{subsec:sample_characterization} and a typical ratio $U/(k_\mathrm{B}T) \approx 10$ from evaporative cooling \footnote{In our case the trap depth is limited along gravity (dotted lines in Fig.~\ref{fig:final_sample}c) due to the residual confinement along the beams.}.
	
	Repeating thermometry measurements over several hours reveal drifts that we incorporate into our temperature estimate as a statistical uncertainty leading finally to $T = \SI{299(21)}{\nano\kelvin}$.
	A similar stability measurement yields a final Li atom number of $N = \num{1.72(15)e+05}$. 

	These quantities, together with the trap model of Sec.~\ref{subsec:sample_characterization}, allow one to calculate the density distributions for both Li and Cs (see Fig.~\ref{fig:final_sample}b and c). For a typical sample used for polaron experiments, we find the local peak Fermi energy $E_\mathrm{F, \, peak} = h \times \SI{37(1)}{\kilo \hertz} = k_\mathrm{B} \times \SI{1.78(5)}{\micro \kelvin}$. The gravitational sag due to the different masses of Li and Cs is less then \SI{5}{\micro \meter}, \textit{cf.} Fig.~\ref{fig:final_sample}c, and reduces the mean Li density sampled by the Cs atoms by \SI{11}{\percent} to about \SI{4.0(7)e12}{\per\cubic\centi\meter} translating to the local Fermi energy sampled by Cs to be $E_F = h \times \SI{32(4)}{\kilo \hertz} = k_\mathrm{B} \times \SI{1.53(17)}{\micro \kelvin}$.
	
	With the known peak densities we checked that both Li-Cs and Cs-Cs inelastic collisions are 
	negligible on typical experimental timescales, as detailed in Appendix~\ref{appendix:inelastic_2b}.
	
	\subsubsection{Magnetic field stability} \label{subsec:magnetic_field_stability}
	
\begin{figure}
		\centering
		\includegraphics[width=1\linewidth]{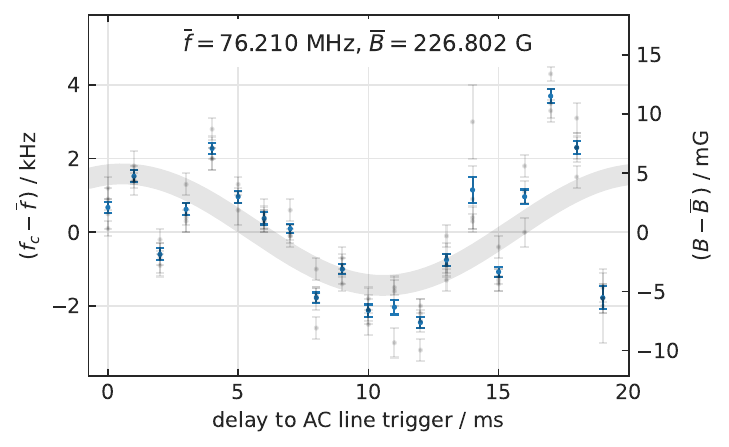}
		\caption{Spectroscopic measurement of the magnetic field as a function of the delay with respect to a trigger that is synchronized to the AC power grid. Each data point (gray) represents the resonance position $f_c$ determined by a fit to a Cs radio-frequency spectrum recorded with \SI{156}{\micro\second} square-shaped $\pi$-pulses. Each spectrum consists of 30 data points and the error bars represent the $1\sigma$-error of the fit. %
			Blue data points are the mean of four independent measurements at each delay with error bars representing their standard deviation. The gray line is a \SI{50}{Hz} sine wave fit serving as a guide to the eye. The deviations of the fitted resonance positions (left $y$-axis) can be converted to deviations from a center magnetic field (right $y$-axis) using the Breit-Rabi formula.}
		\label{fig:line_noise}
	\end{figure}

	Due to the strong dependence of both the impurity-bath interactions and the bare Cs transition frequency on the magnetic field near the \SI{888.6}{\gauss} FR, precise knowledge and control of the magnetic field value are essential for polaron spectroscopy experiments.
	To accurately determine the magnetic field experienced by the atoms, we perform spectroscopy on Li or Cs. 
	When investigating the magnetic field stability on the millisecond time scale by performing Cs spectroscopy, we observe a noticeable 
	dependence of the magnetic field on the \SI{50}{\hertz} frequency of the AC power grid, see 
	Fig.~\ref{fig:line_noise}. This dependence necessitates the synchronization of the experimental sequence to the 50-Hz cycle of the mains power line.
	Due to small drifts in the AC line frequency (up to \SI{40}{\milli\hertz} in 30 minutes), it is not enough to synchronize the start of the $\gtrsim\SI{20}{\second}$ long sequence, but instead we do the synchronization shortly before performing polaron spectroscopy in the final experimental sequence. %
	
\begin{figure}
		\centering
		\includegraphics[width=1\linewidth]{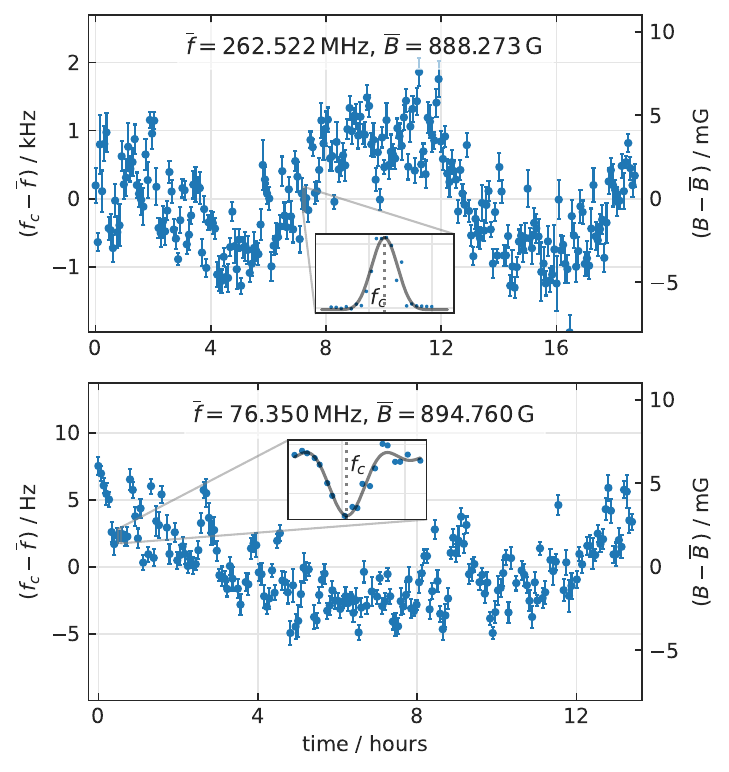}  %
		\caption{Long-term stability of the magnetic field measured with Cs Raman spectroscopy (upper panel) and Li rf spectroscopy (lower panel). Each point represents the resonance position $f_c$ determined by a fit to a spectrum containing 21 data points, see insets, where the vertical error bars represent the estimated standard error of the fit (averaging to \SI{0.8}{\milli \gauss} for Cs and \SI{0.6}{\milli \gauss} for Li). The deviations of the fitted resonance positions (left $y$-axis) can be converted to deviations from a center magnetic field (right $y$-axis) using the Breit-Rabi formula.
			The Cs spectra were recorded using \SI{324}{\micro \second} Blackman-shaped $\pi$-pulses at an effective Rabi frequency of $\Omega_\mathrm{R}=2\pi \times\SI{3.66}{\kilo\hertz}$ leading to a full width at half maximum (FWHM) of the spectrum of \SI{5}{\kilo\hertz} corresponding to \SI{20}{\milli\gauss}. Equivalently, \Li rf square-shaped $\pi$-pulses with a length of \SI{53}{\milli \second} at a Rabi frequency of \SI{9.8}{\hertz} are used, which result in a FWHM of \SI{15}{\hertz} corresponding to \SI{12}{\milli\gauss}. 
		}
		\label{fig:magnetic_field_stability}

	\end{figure}

	On the time scale needed to record one full spectrum ($\sim \num{10}$ minutes), we can quantify the shot-to-shot fluctuations by the minimal spectral linewidth, see insets of Fig.~\ref{fig:magnetic_field_stability} \footnote{Precisely, we vary the pulse length and use the longest one such that the linewidth is Fourier limited.}. 
	Additionally, we observe slow drifts on the time scale of several hours. Since this is also the typical duration of a polaron measurement, we monitor the long-term stability of the magnetic fields by repeatedly performing spectroscopy on either Li or Cs, see Fig.~\ref{fig:magnetic_field_stability}. Taking the spread between the smallest and largest magnetic fields from these measurements as a measure for the statistical uncertainty of the magnetic field value yields \SI{15}{\milli\gauss}.
	This uncertainty of the magnetic field value results in an absolute uncertainty on the interaction parameter $1/(k_\mathrm{F}a)$ of less than $0.03$. Including the systematic uncertainty of the pole of the Feshbach resonance \cite{Ulmanis2015_Universality, Johansen2017_Testing}, the uncertainty on the interaction parameter is around $0.05$. The most important systematic uncertainty on the interaction parameter, clearly dominating the aforementioned contributions, is the uncertainty on $k_\mathrm{F}$ (relative uncertainty of \SI{5.7}{\percent}), which leads to uncertainties on the interaction parameter as large as $0.1$ at $| 1/(k_\mathrm{F}a) | \gtrsim1.5$. %
	
	\section{Cesium Raman spectroscopy} \label{sec:Raman}
	
\begin{figure}
		\centering
		\includegraphics[width=1\linewidth]{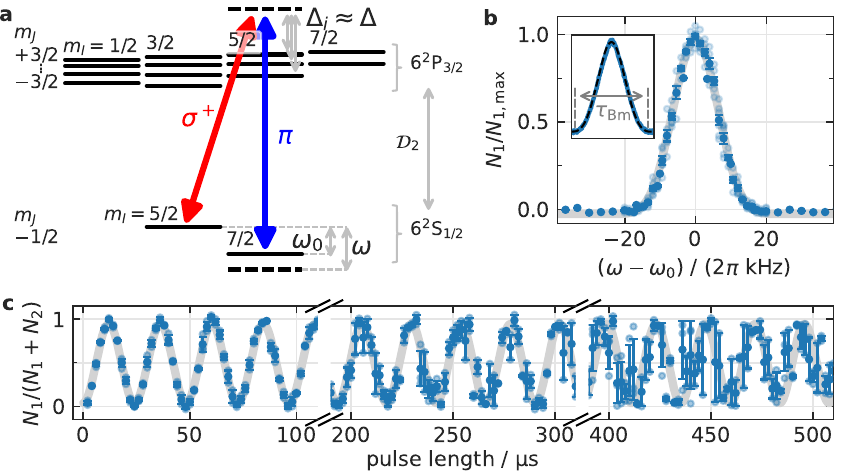} %
		\caption{Cesium Raman spectroscopy. \textbf{a}: The relevant Cs hyperfine levels at $B=\SI{890}{\gauss}$ in the $\ket{m_J, m_I}$ basis. Note that at high magnetic fields the 6$^2$P$_{3/2}$ sublevels are close to the Paschen-Back regime, while the 6$^2$S$_{1/2}$ sublevels are better described in the Zeeman limit. Still, approximately $\ket{2} \approx \ket{m_J=-1/2, m_I=5/2} \approx \ket{F=3, m_F=2}$ and $ \ket{1} \approx \ket{m_J=-1/2, m_I=7/2} \approx \ket{F=3, m_F=3}$ \cite{Steck1998_Cesium} (all calculations use the exact states). Both Raman beams are detuned by $\Delta_i$ from the respective sublevels of the excited state, with the splitting between these levels being much smaller than the detunings. We therefore define the detuning of the Raman light $\Delta \equiv \omega_{\mathrm{R},\sigma^+}- \omega_{\mathrm{D}_2}$ as the difference between the Raman laser frequency and the Cs D$_2$-line. The Raman scheme creates an effective Rabi coupling between the two lowest states with a two-photon detuning $\omega = \omega_{\mathrm{R},\pi} - \omega_{\mathrm{R},\sigma^+}$. \textbf{b}: 
			The resonance as a function of the two-photon detuning. 
			In this case, a Blackman-shaped $\pi$-pulse is used with a pulse length of $\tau_\mathrm{Bm} = \SI{114}{\micro\second}$ %
			as shown by the power–time profile in the inset (blue: measured intensity, black dashed: target pulse shape).
			The measured number of atoms (blue) are in very good agreement with the expected spectrum calculated from the temporal pulse shape (gray line).
			\textbf{c}: Rabi oscillations for $\omega=\omega_0$ with a sinusoidal fit including the data $<\SI{300}{\micro \second}$. The blue data points show the mean and standard deviation across all individual realizations, which are shown with reduced opacity.}
		\label{fig:Raman_all}
	\end{figure}

	To investigate the spectral properties of the heavy Fermi polaron, we need to accurately measure the energy of the transition between the two lowest hyperfine states of the \Cs impurities. To this end, we use a two-photon Raman transition as depicted in Fig.~\ref{fig:Raman_all}a.
	The following paragraphs describe the experimental setup of the Raman spectroscopy, its characterization, and showcase its application as part of the spin-resolved imaging sequence that measures both relevant spin states of Cs to determine the transferred fraction. %
	
	\subsection{Optical and electronic setup} \label{subsec:Raman_setup}
	Raman spectroscopy is implemented by shining two differently polarized light beams on the atoms, as schematically indicated in the upper left panel of Fig.~\ref{fig:Fig1}. In our setup, those two co-propagating beams are derived from the same amplified diode laser \footnote{Toptica TA pro,  max. output power $P_\mathrm{max} \approx \SI{3}{\watt}$} which is freely running at $\Delta = \SI{40}{\giga\hertz}$ %
	blue detuned to the \Cs $D_2$ line (see Fig.~\ref{fig:Raman_all}{a})). A first single-pass acousto-optic modulator (AOM) is used for temporal pulse shaping before the beam is divided into two arms on a polarizing beam splitter. A second AOM, this one in double-pass configuration, is used to shift the light frequency in one arm with respect to the other, thus directly controlling the two-photon detuning $\delta$. The light is subsequently transferred to the main experiment in two separate polarization-maintaining optical fibers, where both beams are finally overlapped with orthogonal polarizations. One of the two polarizations is aligned with the offset magnetic field and consequently couples to the atomic $\pi$ transition, whereas the other beam couples to the $\sigma^\pm$ transitions.
	For the selected scheme, in which the $\sigma^-$ transitions are not involved in the two-photon coupling (\textit{cf.} Fig.~\ref{fig:Raman_all}{a}), the effective two-photon Rabi frequency is given by
	$\Omega_\mathrm{R} = \sum_n \Omega_{\pi,n} \Omega^*_{\sigma^+,n} /(2\Delta_n)$,
	where the sum includes contributions from all excited states $n$ in the $6^2$P$_{3/2}$ manifold \cite{Foot2005_Atomic, Steck2007_Quantum}.%
	
	To ensure reliable spectroscopy, the frequency source of the AOM in double-pass configuration is referenced to a rubidium frequency standard. The radio-frequency control, based on a direct digital synthesis (DDS) frequency synthesizer, allows fast frequency and phase changes, as well as frequency ramps. 
	The electronic control of the single-pass AOM, in turn, enables the generation of arbitrary Raman pulse shapes. Moreover, using a two-frequency driver for this single-pass AOM strongly suppresses thermalization-induced changes in the beam pointing and diffraction efficiency, as described in Ref.~\cite{Frohlich2007_Twofrequency}.
	
	\subsection{Characterization} \label{subsec:Raman_characterization}
	The main purpose of the Raman spectroscopy setup is to probe Fermi polarons, imposing stringent requirements on the spectroscopic resolution. It must be significantly better than the expected polaron signatures, which are on the order of the Fermi energy ($E_\mathrm{F} = h \times \SI{32(4)}{\kilo \hertz}$). To this end, the Raman spectroscopy pulse length is set such that the linewidth is dominated by Fourier broadening, ensuring that additional technical broadening is negligible. %
	To avoid the pronounced side lobes in the spectrum caused by square pulses, we employ pulses with a Blackman temporal shape \cite{Blackman1958_Measurement, Weisstein_Blackman, Steck2007_Quantum}. 
	In this case, the pulse intensity follows the temporal envelope shown in the inset of Fig. \ref{fig:Raman_all}b, and the resulting spectrum, shown in the main panel, demonstrates a purely Fourier-limited response. This temporal pulse shaping leads to a time-dependent Rabi frequency $\Omega_\mathrm{R}(t)$, which results in an effective Rabi frequency $\overline{\Omega} = \int_0^\tau\Omega_\mathrm{R}(t)\mathrm{d}t/\tau$ that is smaller by a factor of $0.42$ than the corresponding Rabi frequency for a rectangular pulse with Rabi frequency $\max_t \Omega_R(t)$.
	
	We characterize the coherence properties of our spectroscopy setup by driving continuous Rabi oscillations (see Fig.~\ref{fig:Raman_all}c).
	For short interrogation times, coherent oscillations are clearly visible, while for longer times noise increases. %
	We attribute this behavior primarily to fluctuations in the laser power, magnetic field noise, and other technical imperfections. Nevertheless, the observed coherence time remains sufficiently long to enable the spectroscopic investigation of the many-body phenomena of interest.
	Ultimately, magnetic field stability limits the spectroscopic resolution. The energy splitting between Cs$\ket{1}$ and Cs$\ket{2}$ changes by $\SI{0.24}{\kilo\hertz\per\milli\gauss}$%
	for small magnetic field changes around an offset field of $\SI{890}{\gauss}$. The magnetic field noise (\textit{cf.} Sec.~\ref{subsec:magnetic_field_stability}) restricts the minimal spectral linewidth to a FWHM of $\SI{5}{\kilo \hertz}$, as shown in the inset of Fig. \ref{fig:magnetic_field_stability}a. This matches the magnetic field resolution that can be achieved using rf spectroscopy on Li (see Fig. \ref{fig:magnetic_field_stability}b). Because the Li transition is much less sensitive to magnetic fields ($\SI{1.3}{\hertz\per\milli\gauss}$ at $\SI{890}{\gauss}$), reaching the same magnetic field resolution requires much longer spectroscopy pulses.
	This spectroscopic resolution is sufficient to resolve the features of the Fermi polaron which are on the order of the Fermi energy $E_\mathrm{F} = h \times \SI{32(4)}{\kilo \hertz}$.
	Furthermore, the Raman spectroscopy setup allows Rabi frequencies up to $\Omega_\mathrm{R} = 2\pi \times \SI{500}{\kilo \hertz}$ at a detuning of $\Delta = \SI{20}{\giga \hertz}$.
	Such a large Rabi frequency is beneficial for robust and fast changes of the populations between two Cs hyperfine states (as applied, \textit{e.g.}, in Sec~\ref{sec:double_imaging}) or driving well-defined operations on the Bloch sphere, even under the presence of interactions, \textit{e.g.}, for Ramsey sequences or spin tomography.
	
	There are certain potential limitations and undesired effects of the Ramsey spectroscopy which need to be considered.
	The near-resonant light, which the Raman process is based on, creates atom loss and decoherence due to single photon scattering. But since both of these processes are suppressed with respect to the Rabi frequency by $\Gamma_\mathrm{D2}/\Delta \approx 10^{-3}$ where $\Gamma_\mathrm{D_2}$ is the natural line width of the \Cs D$_2$ transition, they can be neglected in our case, as shown in more detail in Appendix \ref{appendix:Raman_photon_scattering}. The trapping potential induced by the Raman beams is significant, but due to their large beam waist of $w_\mathrm{Raman} \approx \SI{750}{\micro \meter}$, which is 15 times larger than the one of the trapping potential, the resulting dipole forces are negligible. %
	For the weak intensities used for polaron spectroscopy, the influence of the Raman beams on the considered \Li-\Cs Feshbach resonance can also be safely ignored. In principle, differential AC Stark shifts on the states involved in the Feshbach resonance can shift its center \cite{Lous2018_Probing, Bauer2009_Combination}, but measurements confirm that this effect is negligible, as detailed in Appendix~\ref{appendix:FR_shift}.
	
	\subsection{Measuring transferred fraction: state-resolved Cesium imaging} \label{sec:double_imaging}

	To suppress the influence of fluctuations of the total number of Cs atoms on the determination of the transferred fraction between states Cs$\ket{1}$ and Cs$\ket{2}$, it is advantageous to measure the number of atoms in both hyperfine states independently using conventional absorption imaging \cite{Ketterle_Making}. Absorption images are recorded using imaging optics with a numerical aperture of $\mathrm{NA}=\num{0.15}$ and a Rayleigh resolution limit of \SI{9}{\micro\meter}. With $N_i$ the number of atoms in state Cs$\ket{i}$, the transferred fraction can then be calculated as $N_1/(N_1+N_2)$. %

	At offset fields around \SI{900}{\gauss}, the Cs$\ket{1}$ state can be directly imaged after optically pumping the atoms to the $\ket{F=4,m_F=4}$ state, from which the closed imaging transition $\ket{F=4,m_F=4}\leftrightarrow\ket{m_I^\prime=7/2,m_J^\prime=3/2}$ can be driven, that leaves the Cs$\ket{2}$ state unaffected \cite{Steck1998_Cesium, Berninger2011_Universal}. The Cs level structure does not provide an equivalent closed transition for direct imaging of the Cs$\ket{2}$ state. Therefore, we use the Raman laser coupling (see Fig.~\ref{fig:Raman_all}) to exchange the populations of the two lowest hyperfine states before performing a second imaging sequence. %
	Prior to the state transfer, Li is removed to avoid interspecies scattering during the imaging sequence. For a reliable Cs$\ket{1}\leftrightarrow\mathrm{Cs}\ket{2}$ transfer, a fast and magnetic-field-insensitive method is preferred. For this we use either a resonant $\pi$ pulse ($\Omega_\mathrm{R} = 2\pi \times (\num{180}...\num{500})\;\si{\kilo \hertz}$) or an adiabatic rapid passage (sweeping the frequency in a linear ramp from $-\Delta f$ to $+\Delta f \approx \SI{350}{\kilo \hertz}$ around the transition frequency within $t \approx \SI{0.5}{\milli \second}$ at a Rabi frequency of $\Omega_\mathrm{R} \approx 2\pi \times \SI{40}{\kilo \hertz}$). %
	
\begin{figure}
		\centering
		\includegraphics[width=0.9\linewidth]{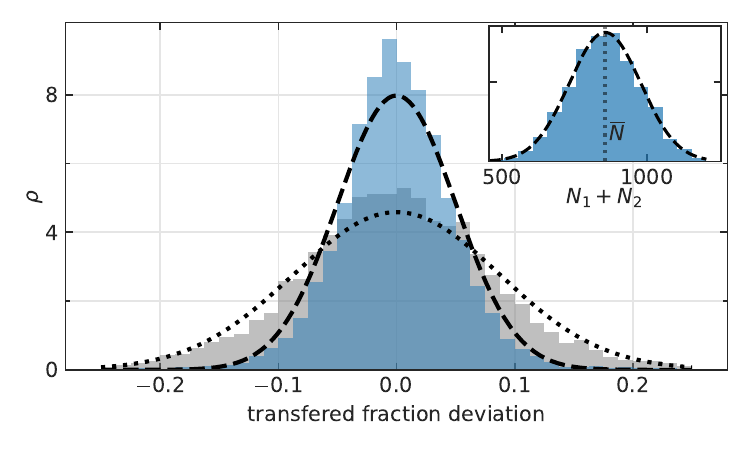}
		\caption{Characterization of Cs imaging at $\SI{889}{\gauss}$ in the absence of Li.%
			We show the normalized histogram of the measured transfer fraction, expressed as a deviation from its expected mean value. We compare two cases: in gray the transferred fraction is extracted from imaging only the Cs$\ket{1}$ state, while in blue the transferred fraction is deduced from considering images of the Cs$\ket{1}$ and Cs$\ket{2}$ states. Details on the measurement protocol are given in Appendix~\ref{appendix:doubleimg}. The dotted (dashed) lines show the corresponding Gaussian distributions with a standard deviation of $0.087$ ($0.050$). The inset shows a histogram of the total cesium atom number obtained from more than 2000 %
			measurements, with the dashed line indicating a Gaussian distribution with mean $\overline{N} = 856$ and standard deviation $\sigma_N = 123$.
		}
		\label{fig:CsDoubleImg}
	\end{figure}

	Measurements in the absence of Li reveal that, independent of the applied method, we suffer from roughly the same finite state-transfer efficiency. This necessitates a calibration of the detected atom number in the second image, which is done as described in Appendix~\ref{appendix:doubleimg}.
	After performing this calibration, we quantify the improvements by measuring the transferred fraction as described above compared to considering only the Cs$\ket{1}$ state population. To facilitate this comparison, we introduce the pseudo-transferred fraction, defined as the population in the Cs$\ket{1}$ state normalized to the estimated total Cs population. Figure~\ref{fig:CsDoubleImg} shows histograms of the deviations of the two transferred fractions from their respective expected mean values. Indeed, measuring the population of both Cs spin states instead of acquiring only one, mitigates fluctuations in the total atom number (see inset) and thus reduces 
	measurement noise by a factor of $\approx 1.7$. %

	\section{Spectroscopy of a heavy Fermi polaron} \label{sec:spectra}
	
	With all ingredients in place, we demonstrate the creation of a Fermi polaron by recording the impurity energy spectrum as a function of impurity-bath interactions. To this end, we employ injection spectroscopy, starting from an effectively non-interacting reference state and transferring the impurity into the interacting polaron state.
	We tune the Li$\ket{2}$-Cs$\ket{1}$ interactions, while Li$\ket{2}$-Cs$\ket{2}$ is only weakly interacting with approximately constant $a_\mathrm{Li\ket{2}-Cs\ket{2}} \approx \SI{-41}{\bohr}$ (\textit{cf.} Fig.~\ref{fig:FB_resonances}), making Cs$\ket{2}$ an excellence reference state. To perform spectroscopy, we prepare the sample in the weakly interacting Li$\ket{2}$-Cs$\ket{2}$ channel as described in Sec.~\ref{sec:sequence} and then use Raman spectroscopy to drive Cs$\ket{2} \rightarrow$ Cs$\ket{1}$ (see Fig~\ref{fig:Fig1}). The transferred fraction is subsequently determined by spin-resolved absorption imaging, as outlined in Sec.~\ref{sec:double_imaging}. To avoid side lobes in the frequency spectrum, we use a Blackman-shaped temporal envelope for the spectroscopy pulse (see Sec. \ref{subsec:Raman_characterization}). To maximize the spectral resolution given our magnetic field stability, we choose a pulse duration of \SI{372}{\micro\second}. The spectroscopy power is adjusted to  realize a $\pi$-pulse on the bare Cs $\ket{1}\leftrightarrow\ket{2}$ transition (corresponding to a %
	bare Rabi frequency of $\Omega_\mathrm{R}= 2\pi \times \SI{3.2}{\kilo\hertz}$). %
	At each interaction strength, controlled by the magnetic field, we also measure a reference spectrum without the Li Fermi sea by removing the Li atoms prior to performing spectroscopy with a resonant light pulse. From this reference spectrum we obtain the bare Cs resonance frequency and an absolute magnetic field calibration. These quantities then allow us can calculate the Li$\ket{2}$-Cs$\ket{1}$ scattering length $a \equiv a_\mathrm{Li\ket{2}-Cs\ket{1}}$ and ultimately the interaction parameter $1/(k_\mathrm{F}a)$. Furthermore, we subtract the bare resonance frequency from each transition frequency in the injection spectra.
	
\begin{figure}[t!]
		\centering
		\includegraphics[width=1\linewidth]{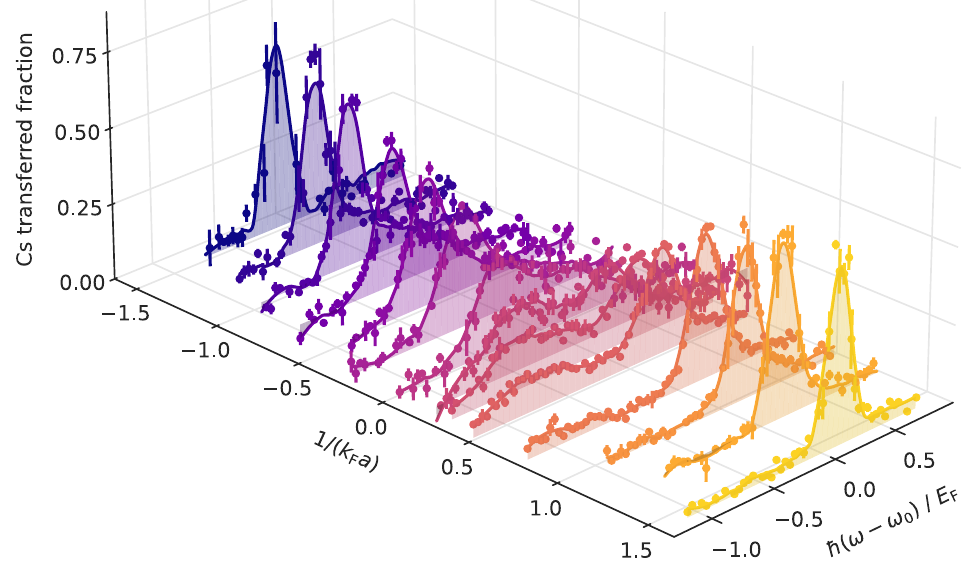}
		\caption{Injection spectra at different interaction strengths parametrized as the inverse of the product of the Fermi momentum $k_F$ and the scattering length $a$ (coded in different colors). The spectral response is measured as the fraction of atoms transferred to the final state Cs$\ket{1}$. Data points are the average of at least three measurements, with error bars denoting the standard error of the mean. The solid line is a guide to the eye (computed as a Savitzky-Golay filter of order 2 with a window of $0.6E_\mathrm{F}$). %
		}
		\label{fig:injection_3d}
	\end{figure}

	Fig.~\ref{fig:injection_3d} summarizes the measured transferred fractions obtained for different interaction strengths (in units of the Fermi momentum) with the spectroscopy settings as outlined above. For small negative scattering lengths $a$, the attractive polaron appears as a peak centered at negative energies. With increasing interaction strength, the spectrum broadens and its center shifts to more negative energies. For small positive scattering lengths, the repulsive polaron appears as a similar peak at positive energies, which broadens and shifts towards higher energies as the resonance is approached. Close to unitarity, both features are visible, giving rise to a more complex spectral structure. Detailed analysis of these injection spectra, along with comparisons with theoretical models, will be published elsewhere \cite{KromRautenberg2026_spectroscopy}.

\section{Conclusion}

We have described in detail an experimental platform for creating and detecting Fermi polarons in a strongly mass-imbalanced quantum-gas mixture consisting of \Cs impurities implanted into a Fermi sea of \Li atoms. The system enables the reproducible preparation of a deeply degenerate lithium Fermi gas with well-characterized density and temperature in a single optical trap together with cesium impurities. Importantly, the Cs atoms populate a region of almost constant fermion density. Precise magnetic-field control allows for tunable interspecies interactions close to a broad Li-Cs Feshbach resonance. The availability of an additional impurity state that remains essentially noninteracting with the bath at the same magnetic field provides an ideal reference state for spectroscopy of the polaron. Optical Raman spectroscopy combined with state-resolved detection of both cesium hyperfine states in a single experimental realization provides sensitivity sufficient to resolve subtle many-body effects in the spectra.  Strong coupling between the two Cs states yields Rabi frequencies $\Omega_\mathrm{R} \gg E_\mathrm{F}/\hbar$. Overall, the combination of high stability, controllability, and detection fidelity establishes a versatile platform for studying impurity physics in ultracold quantum gases. In particular, we presented well-resolved injection spectra of a Fermi polaron for both attractive and repulsive interactions. 

By letting the polaron thermalize after initial excitation, ejection spectroscopy \cite{Parish2025_Fermi, Massignan2026_Polarons} can be implemented, enabling a detailed comparison of spectral response functions and their thermodynamic interpretation \cite{Liu2020_RadioFrequency, Liu2020_Theory}. Such detailed spectra can serve as a benchmark for differentiating different theoretical models \cite{Hu2022_Fermi, Hu2024_Spectral, Adlong2020_Quasiparticle, Liu2019_Variational, Chen2025_MassGap, Drescher2024_Bosonic} and thus have the potential to contribute to further clarifying the nature of the heavy Fermi polaron.
Time-domain spectroscopy protocols such as Ramsey and spin-echo sequences \cite{Knap2012_TimeDependent} provide complementary information to be compared to spectral functions deduced from spectroscopy. This opens detailed access to different stages in the temporal evolution of the polaron, ranging from its initial formation dynamics to thermal dephasing processes \cite{Cetina2016_Ultrafast, Cetina2015_Decoherence, Skou2021_Nonequilibrium, Skou2022_Life}. Well-controlled, continuous driving of the impurity between interacting and noninteracting states provides access to the dynamical response of the driven Fermi polaron %
\cite{Vivanco2025_strongly} with possible signatures of the Anderson Orthogonality Catastrophe \cite{Adlong2021_Signatures, Schmidt2018_Universal}. The variety of these different approaches enables a conclusive study of the Fermi polaron in the limit of a very heavy impurity. By accessing the extreme mass-imbalance regime offered by the \Li–\Cs mixture, new investigations of impurity physics and many-body phenomena beyond previously explored parameter ranges become possible.

\section{Acknowledgments}
We thank R.~Grimm, C.~Baroni, E.~Dobler, G.~Roati, M.~Zaccanti, F.~Scazza, C.~Chin, S.~Jochim, and G.~Zürn for technical discussions and our theory collaborators E.~Dizer, O.~Bleu, M.~Drescher, R.~Schmidt, and T.~Enss. 
We further thank K.~Welz, R.~Freund, A.~Schürg, and F.~Borchers for their work in the early stages of the experiment. %
This work is funded by the Deutsche Forschungsgemeinschaft DFG (German Research Foundation) under Project-ID 273811115 – SFB 1225 ISOQUANT, and the Heidelberg Excellence Cluster STRUCTURES (EXC 2181/1 - 390900948). Support by the Heidelberg Center for Quantum Dynamics is gratefully acknowledged. M.R., T.K, and E.L. acknowledge support by the International Max Planck Research School for Quantum Dynamics in Physics, Chemistry and Biology (IMPRS-QD). M.R. acknowledges financial support from the German Academic Exchange Service (DAAD).

\FloatBarrier
\appendix

\section{Lithium thermometry} \label{appendix:Li_thermometry}

\begin{figure}
	\centering
	\includegraphics[width=1\linewidth]{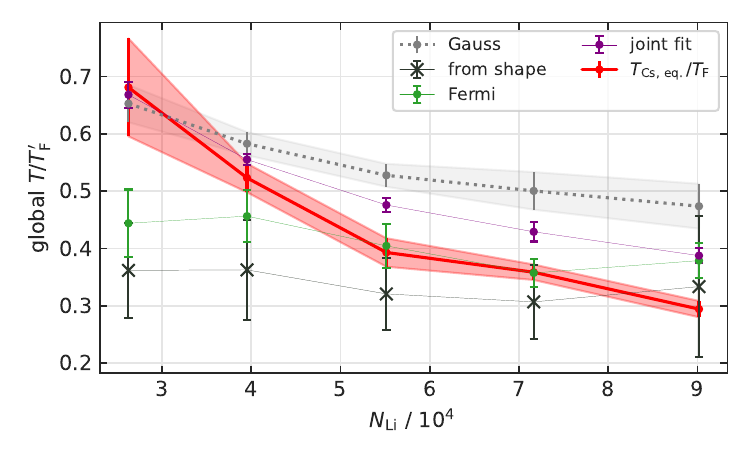}
	\caption{Comparison of different methods to determine the reduced temperature $T/T_\mathrm{F}^\prime$ of the Li Fermi sea. Data points are averaged over different Li-Cs interaction times between \SIrange{50}{700}{\milli\second} as well as different values of time-of-flight expansion times, where applicable. Error bars are standard deviations of the data before averaging, except for the red points where error bars are dominated by the fit uncertainty of the Cs equilibrium temperature.}
	\label{fig:thermometry}
\end{figure}

Accurately determining the temperature of a degenerate Fermi gas from absorption pictures is challenging. To demonstrate the difficulties, we briefly discuss different thermometry methods using a set of time-of-flight absorption images taken with both Li and Cs as a function of the time that they are interacting \footnote{during their ``interaction time'', Li and Cs are moderately interacting with a Li$\ket{2}$-Cs$\ket{1}$ $s$-wave scattering length of $a_\mathrm{Li-Cs} = \SI{250}{\bohr}$ while Cs is only very weakly interacting among itself with $a_\mathrm{Cs-Cs} < \SI{50}{\bohr}$. Before imaging, the magnetic field is adjusted such that Li-Cs are only weakly interacting with $a_\mathrm{Li-Cs} \approx \SI{45}{\bohr}$ while Cs is attractively interacting with $a_\mathrm{Cs-Cs} \approx \SI{-400}{\bohr}$}. Additionally, we change another parameter, which mainly reduces the Li atom number, thereby increasing $T/T_\mathrm{F}^\prime$.

There are two distinct ways to determine the reduced temperature $T/T_\mathrm{F}^\prime$ from Li absorption images. As detailed in \cite{Ketterle2008_Making}, either from directly fitting the shape of the two-dimensional density distribution or calculating the global $T/T_\mathrm{F}^\prime$ from the measured temperature $T$ and atom number $N$ along with the known harmonically averaged trap frequencies $\overline{\omega}$ as
\begin{equation} \label{eq:TF_global}
	k_\mathrm{B} T_\mathrm{F}^\prime = \hbar \overline{\omega} \left(6N\right)^{1/3} \quad .
\end{equation} 

Already determining a temperature from a series of absorption pictures as a function of the time-of-flight $t$ requires fitting the two-dimensional density distributions to determine the cloud sizes at times $t$. A thermal cloud is well described by a two-dimensional Gaussian function with cloud radii $\sigma_i(t) = \sqrt{{k_BT}/{(m\omega_i^2})} \times b_i(t)$, whereas in the degenerate regime, the integrated density of a non-interacting Fermi gas follows
\begin{equation} \label{eq:n2d_Fermi}
	n_{2d}(x, y) = n_{2d,0} \frac{\mathrm{Li_2}\left(-\exp\left[q-\left(x^2/r_x^2 + y^2/r_y^2 \right) f\left(e^q\right)\right]\right)}{\mathrm{Li_2}\left(-e^q\right)}
\end{equation}
with radii $r_i(t) = \sqrt{{k_BT}/({m\omega_i^2}) \ f\left(e^{q}\right)} \times b_i(t)$. Here, $q = \mu/(k_BT)$ is the logarithm of the fugacity and determines the shape of the cloud and $f(x) = (1+x)/x\ln(1+x)$. 
Both $\sigma_i$ and $r_i$ expand with the same time-dependence, which in the absence of external potentials is given by $b_i(t) = \sqrt{1+\omega_i^2 t^2}$. For the following analysis, the additional effects of the saddle potential from the magnetic field curvature on the expansion of the cloud have been taken into account, although these effects are negligible for the comparably short times-of-flight $t<\SI{3.5}{\milli\second}$. 

In Fig.~\ref{fig:thermometry} we compare the results of the following methods to extract the global $T/T_\mathrm{F}^\prime$. 
The red data points mark our best estimate for $T/T_\mathrm{F}^\prime$ and are calculated using the atom number $N$ from two-dimensional Gaussian or Fermi fits \footnote{global $T_\mathrm{F}$ calculated from atoms numbers from two-dimensional Gaussian fits and those determined by fitting Eq.~(\ref{eq:n2d_Fermi}) agree within \SI{2}{\percent}} as well as the Cs equilibrium temperature, determined from an exponentially decaying fit to the time evolution of the Cs temperature as a function of the interaction time between Li and Cs, as described in Sec.~\ref{subsec:thermometry}, \textit{cf.} Fig.~\ref{fig:final_sample}d. The red shaded area marks the combined uncertainties from the error of the fit to the Cs temperature evolution as well as the statistical uncertainties on the atom number.

For the gray data points connected by the dotted line, we extract $N$ and $\sigma_i(t)$ from two-dimensional Gaussian fits to the \Li cloud. The temperature $T$ is then obtained by fitting the temporal evolution of $\sigma_i(t)$, and the reduced temperature $T/T_\mathrm{F}^\prime$ is calculated using Eq.~(\ref{eq:TF_global}).
This method should work well for thermal or weakly degenerate samples and break down at increasing degeneracy. 
Alternatively, for the green data points, we can use Eq.~(\ref{eq:n2d_Fermi}) to determine $N$ and extract the temperature from fitting the time dependence of $r_i(t)$ and again compute $T/T_\mathrm{F}^\prime$ using Eq.~(\ref{eq:TF_global}). 
This method is expected to be consistent with extracting the reduced temperature
\begin{equation}
	T/T_\mathrm{F}^\prime = \left[-6\mathrm{Li}_3(-e^q)\right]^{-1/3}
\end{equation}
directly from the two-dimensional fitting parameter $q$ in Eq.~(\ref{eq:n2d_Fermi}). The results of the latter procedure are shown as dark green data points and do surprisingly not agree with the previously mentioned temperature determination (green data points).

To enforce consistency, we can apply a joint fitting routine that fits a global $q$ parameter for a complete sequence of time-of-flight pictures at different times $t$, shown in purple. This fitting routine takes $T_\mathrm{F}^\prime$ as an additional input parameter which is again calculated from Eq.~(\ref{eq:TF_global}) using $N$ from independent two-dimensional fits and known trap frequencies.

In conclusion, we observe that while the Gauss fits agree well for high temperatures, both methods depending on fitting the shape of the fermionic cloud do not adequately capture the change in degeneracy across the data set. While the joint fitting routine improves on this and delivers qualitatively convincing results, the absolute temperatures are still not adequately reproduced. We attribute these deviations mainly to the difficulties in resolving small changes in cloud shapes with our imaging system, which is designed to provide the best resolution at both Li and Cs wavelengths. Because of this compromise in imaging resolution, we cannot resolve the shape of the Li cloud well enough to reliably extract the temperature directly from absorption images. Usual difficulties, including a possible detuning from the atomic resonance frequency, motional blurring, imaging imperfections, etc. further complicate direct thermometry. 

Due to its conceptual simplicity and robustness, using Cs as a thermometer is considered the most reliable method which is therefor used to determine the final Li temperatures given in the main text. 

\section{Inelastic collisions} \label{appendix:inelastic_2b}

Since preparation and spectroscopy require working with excited internal states of both Li and Cs, inelastic two-body scattering is generally a concern. In the following, we estimate scattering- and loss timescales relevant for our polaron experiments.

To estimate the expected elastic and inelastic rate coefficients, we assume that collisions are dominated by their $s$-wave contribution and that the thermal average can be approximated by evaluating the cross-sections at a single collision energy corresponding to the trap temperature \cite{Bohn1997}. The corresponding two-body collisional rates are obtained by multiplying the rate coefficients by the atomic density.
For the following estimates we use typical densities of $n_\mathrm{Li} = \SI{4e12}{\per\centi\meter\cubed}$ for Li and $n_\mathrm{Cs} = \SI{8e11}{\per\centi\meter\cubed}$ for Cs and use the rate coefficients calculated at \SI{888.6}{\gauss}.

Since Li$\ket{2}$-Cs$\ket{1}$ is stable against two-body collisions (the Cs level splitting at typical working fields is $\sim\SI{260}{\mega \hertz}$ and thereby much larger than the Li splitting of $\sim\SI{80}{\mega\hertz}$), our only concern is scattering in the Li$\ket{2}$-Cs$\ket{2}$ channel.
With the typical densities mentioned above, we find an inelastic two-body scattering timescale of $\tau_{2,\,\mathrm{inel}} > \SI{5}{\second}$ \footnote{The cross-sections are obtained from full coupled-channel calculations performed at a collision energy corresponding to $\SI{1}{\micro\kelvin}$ by Arthur Christianen through private communication, using the interaction potentials determined by Pires \textit{et al.}~\cite{Pires2014_Observation}. Numerical values can be found in \cite{Lippi2024_phd}.}. %

Another concern are spin-changing two-body collisions between two Cs$\ket{2}$ atoms. Here we find a collisional timescale of $\tau_{2,\, \mathrm{inel,\,Cs2Cs2}} \approx \SI{70}{\milli\second}$. %
For the inelastic two-body collisions between Cs$\ket{1}$ and Cs$\ket{2}$ atom, which are only relevant during the Raman pulses or for more complicated schemes such as Ramsey interferometry or continuously driven Rabi oscillations, we find very long timescales of several hundred seconds 
\footnote{The inelastic rate coefficients are obtained from full coupled-channel calculations performed at a collision energy corresponding to $\SI{1}{\nano\kelvin}$. The results for the Cs$\ket{2}$-Cs$\ket{2}$ channel were taken from Frye \textit{et al.}~\cite{Frye2019a}, whereas those for the Cs$\ket{1}$-Cs$\ket{2}$ channel were provided through private communications with Matthew Frye and Jeremy Hutson. Numerical values can be found in \cite{Lippi2024_phd}.}. 

For inelastic three-body collisions between three Cs$\ket{2}$ atoms, we find a loss timescale on the same order as $\tau_{2,\, \mathrm{inel,\,Cs2Cs2}}$ \cite{Dincao2005_Scattering}. A systematic study would be needed to estimate the different three-body loss coefficients \cite{Horvath2024_BoseEinstein}, since their calculation requires detailed knowledge of the underlying interaction potentials \cite{Dincao2004_Limits}. Experimentally, we do not observe fast Cs$\ket{2}$ losses on the timescale of a few milliseconds that would limit polaron experiments. 

Overall, these moderate losses concerning the Cs$\ket{2}$ atoms play a minor role during the sample preparation and, most importantly, do not limit our polaron experiments, where Cs atoms are only in the Cs$\ket{2}$ state for a few milliseconds before spectroscopy.

\section{Photon scattering during the Raman process}
\label{appendix:Raman_photon_scattering}
\begin{figure}
	\centering
	\includegraphics[width=\linewidth]{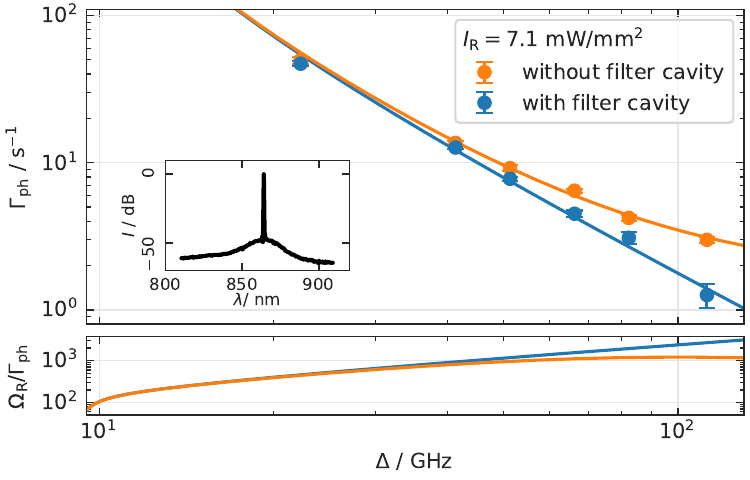}
	\caption{Photon scattering losses in a Raman process for different detunings. The experimental atom number reduction with Raman pulse length is mapped to a photon scattering rate as explained in the main text and shown in the upper graphs, where the error bars consider only the estimated standard error of the measured atom number reduction. As shown in the inset, the used diode laser suppresses its amplification background $\approx\SI{50}{\dB}$ %
		\cite{TAtoptica}, but since a small intensity remains on the resonance frequency, a difference between the orange data (no background suppression) and the blue data with an additional background suppression is visible. The blue line is the theoretical expected scattering for certain detuning, leaving the intensity as a free parameter to take Raman beam misalignment on the atomic cloud and model limitation into account ($I_\mathrm{fit}/I_\mathrm{R} = \num{0.33(1)}$). For the orange line an additional offset in the photon scattering rate of $\Gamma_\mathrm{ph} = \SI{1.7(2)}{\per\second}$ is optimized to the data. The lower panel compares the Rabi frequency with the photon scattering rate for equal beam intensity.}
	\label{fig:raman_losses}
\end{figure}

An intrinsic side effect of Raman spectroscopy is parasitic resonant scattering. To quantify this photon scattering rate, the Raman beam is directed onto the cloud, and the remaining atom number is measured after a waiting time for different pulse durations. The observed loss is mapped to a scattering rate using a simplified model: (i) the photon scattering heats up the cloud, and after the Raman pulse, the system is described by a Boltzmann distribution with a higher temperature; (ii) this new Boltzmann distribution is truncated at the trap depth, leading to atom losses; (iii) additionally, evaporative losses during a rethermalization to the initial temperature are taken into account. %
This model is reasonable, since the recoil energy \Cs $E_{rec}$ is significantly smaller than the trap depth, such that a photon scattering event never leads to a direct atom loss, and a homogeneous modeling - as described in (i) - is more appropriate. Additionally, the thermalization timescale, which is experimentally determined to be on the order of several hundred milliseconds, is typically significantly longer than Raman pulse lengths, which justifies the truncation losses and the stepwise model in general. %
A comparison of the exponential-like reduction of atom number with Raman pulse length in the experiment and the exponential-like reduction of atom number with increasing photon scattering events in the model allows us to compare the experimental losses with the one predicted by theory for a given light intensity and detuning. In Fig.~\ref{fig:raman_losses} the losses are investigated in the case that the amplified diode laser is used directly and additionally when its broad amplification background (see inset in Fig.~\ref{fig:raman_losses}) is filtered out by a cavity (finesse of around 150). %

As shown in the lower panel of Fig.~\ref{fig:raman_losses}, the photon scattering rate can be safely ignored. Also, additional filtering with a cavity becomes only relevant for detuning larger \SI{50}{\giga\hertz} and is therefore not done in the presented results of the impurity spectroscopy.

\section{Shift of the Feshbach resonance center due to Raman light} \label{appendix:FR_shift}

Since we use Raman spectroscopy on Cs to map out the polaron energies, we need to make sure that the spectroscopy does not distort the measurement outcome. One concern when using near-resonant light close to a magnetic Feshbach resonance is that the Raman light can induce differential light shifts in the states involved in the Feshbach resonance and thereby effectively shift the pole of the resonance, \textit{cf.} \cite{Jag2014_Observation, Lous2018_Probing, Bauer2009_Control, Bauer2009_Combination}. Since the $B_0=\SI{888.577(10)(10)}{\gauss}$ Li$\ket{2}$-Cs$\ket{1}$ Feshbach resonance considered here with a resonance strength of  $s_\mathrm{res} = 0.66$ is a broad resonance \cite{Johansen2017_Testing}, the effect of the Raman light on the resonance position is expected to be small. To make sure that there is indeed no significant shift for the polaron spectroscopy, we measure the shift experimentally at much stronger Raman intensities than used later. Additionally, we keep the Raman detuning $\Delta = \SI{41.4}{\giga \hertz}$ %
at a comparable value as in the spectroscopy experiments.

\begin{figure}
	\centering
	\includegraphics[width=1\linewidth]{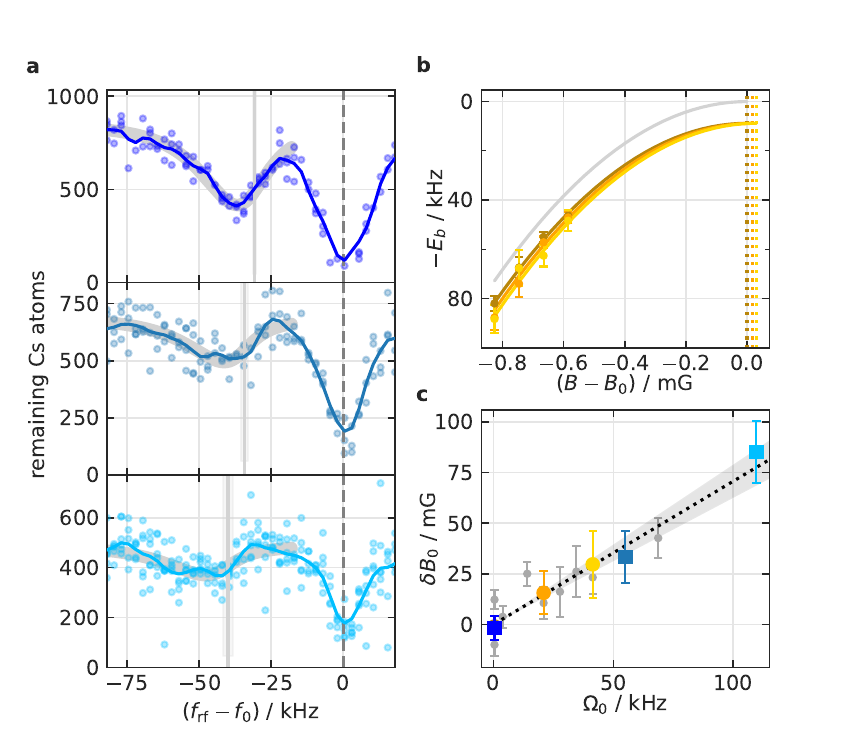}
	\caption{Light-induced shift of the Feshbach resonance center. In \textbf{a} the remaining Cs atoms due to strong losses following dimer association via Li rf spectroscopy are shown, recorded at a magnetic field of \SI{888.0986+-0.0022}{\gauss}. The fitted line-shape models allow us to extract the bound state energies (gray vertical lines; shaded regions indicate the fit uncertainties), which are used to calculate the resonance positions shown in panel \textbf{c} with corresponding colored squares. The upper graph is recorded without Raman light, the one in the middle with a Raman intensity corresponding to a bare Cs Rabi frequency of $2 \pi \times \SI{55.1+-0.4}{\kilo Hz}$ %
		and the lower to $2 \pi \times \SI{109.7+-0.9}{\kilo Hz}$. %
		In \textbf{b} the measured bound state energies $E_b$ at different magnetic fields $B$ (solid lines) for different Raman intensities (color coded; brown: no Raman light, orange: bare Cs Rabi frequency $\omega = 2\pi \times \SI{20.89+-0.20}{\kilo\hertz}$, yellow: $\omega = 2\pi \times \SI{41.39+-0.34}{\kilo\hertz}$) allow the determination of the FR pole (dotted vertical lines in the same colors). The gray curve shows the theoretical dimer binding energies without Raman light and without a Li Fermi sea. The extracted resonance shift is shown with the correspondingly colored circles in panel \textbf{c} .
		Figure \textbf{c} summarizes the finding on the resonance shift due to the Raman light. Its shift is correlated with the Raman light intensity, which is measured in units of bare Rabi frequency at a detuning of $\Delta = \SI{41.4}{\giga \hertz}$. A linear fit extracts a slope of $\SI{0.71(8)}{\milli \gauss \per \kilo \hertz}$.} 
	\label{fig:RamanFBshift_all}
\end{figure}

To determine the pole of the Feshbach resonance of interest, we map out the energy of the universal LiCs dimer state at different magnetic fields on the repulsive side of the Feshbach resonance, cf. \cite{Ulmanis2015_Universality}. As we need to do this independently of the Raman laser, we use conventional rf spectroscopy on the Li atoms in the following way: Starting from the non-resonant channel, in our case the Li$\ket{1}$-Cs$\ket{1}$ state, we apply an rf pulse that, when resonant with the dimer state, transfers part of the Li atoms from Li$\ket{1}$ into the Li$\ket{2}$ state. These atoms are subsequently lost in rapid dimer+Li losses, which are observable as an increased loss on the remaining Cs atoms. Repeating this procedure while additionally shining in the Raman lasers (with off-resonant two-photon detuning $(\delta - \delta_0)/\Omega_0 \gg 1$, \textit{i.e.} without altering the Cs spin state populations), we can extract how much the bound state position shifts as a function of the Raman laser intensity. A line-shape model taken from \cite{Ulmanis2015_Universality} is used to extract the bound state energy of each loss resonance, as exemplified in Fig.~\ref{fig:RamanFBshift_all}a.

The bound state energy $E_b$ at a given magnetic field $B$ provides direct access to the Feshbach resonance position $B_0$ using
\begin{equation}
	E_b =  \frac{\hbar^2}{2\mu\left(a - \bar{a} + R^{*}\right)^2} \ ,
\end{equation}
with $\mu$ the \Li-\Cs reduced mass, $a(B) = a_{bg} - a_{bg}\Delta B / \left(B-B_0\right) $ the Li$\ket{2}$-Cs$\ket{1}$ $s$-wave scattering length, $\overline{a}=0.955 978 . . . R_{vdW} \approx \SI{43}{\bohr}$ the mean scattering length of the Li-Cs van der Waals potential, the background scattering length $a_{bg} = \SI{-29.6}{\bohr}$ \cite{Johansen2017_Feshbach}, %
and the remaining parameters of the resonance as given at the end of Sec.~\ref{sec:sequence}.

Since our initial state is not a thermal ensemble of Li atoms but a degenerate Fermi sea with $T/T_\mathrm{F} \approx 0.4$ a shift with respect to the dimer state of up to the Fermi energy of up to $E_\mathrm{F} \approx h \times \SI{30}{kHz}$ %
is expected \cite{Mora2009_Ground, Punk2009_Polarontomolecule}. %
For the moderate impurity–bath interaction strengths considered here, the shift is assumed to be independent of the interaction strength over the probed range. To account for this effect, a constant offset is subtracted from our data, as indicated in Fig.~\ref{fig:RamanFBshift_all}{b}.

While the determination of the absolute position of the Feshbach resonance strongly depends on this model parameter, its relative shift $\delta B_0 (\Omega_0)$ induced by the Raman light is numerically robust against this offset. Thus, the correlation between the induced shift and the Raman intensity can be probed as shown in Fig.~\ref{fig:RamanFBshift_all}{c}.
Since only values up to $\Omega_0 = 2\pi \times \SI{3.2}{\kilo \hertz}$ %
were used in the spectroscopy experiments presented above, this shift is significantly smaller than our experimental uncertainty on the magnetic field (see Sec.~\ref{subsec:magnetic_field_stability}) and can therefore be safely neglected.

\section{Calibration of the imaging parameters}
\label{appendix:doubleimg}
In the case of coherently driven oscillations between the two lowest hyperfine states of Cs, one notices that the detected population in the Cs$\ket{2}$ state does not exactly mirror the population detected in the Cs$\ket{1}$ state. To clarify this behavior, we consider several %
large datasets of coherent oscillations of Cs in the absence of Li measured at an offset field of $\SI{889}{\gauss}$. %

We start from a sample as described in the main text in Sec.~\ref{sec:sequence}, but prior to the final spectroscopy pulse after reaching the target field (\circled{8} in Fig.~\ref{fig:FB_resonances}) the Li sample gets entirely removed to avoid any influence of the Li-Cs interaction. As briefly mentioned in paragraph~\ref{subsec:sample_preparation} and more detailed in \cite{Lippi2024_Experimental} the initial population of the Cs hyperfine states is set by the Raman-sideband cooling protocol, which prepares the vast majority of the sample in the Cs$\ket{1}$ state. %
The remaining Cs atoms populate the higher Cs states, with a strongly decreasing probability. Because the trapping potential of the Cs$\ket{3}$ and higher-lying hyperfine states vanishes during the subsequent levitation, their populations are expected to be negligible. Accordingly, the following treatment can be restricted to the two lowest-lying hyperfine states of Cs. After repeated changes in the offset magnetic field and waiting times of several seconds during the preparation sequence, the Cs sample can be described as a mixed state of these two states, with a small but finite population in the undesired state. During the final magnetic-field ramp (\circled{8} in Fig.~\ref{fig:FB_resonances}), Li$\ket{2}$--Cs$\ket{1}$ three-body losses close to their FB further reduce this parasitic population to only a few percent of the total Cs atom number.%

\begin{figure}[t]
	\centering
	\includegraphics[width=0.8\linewidth]{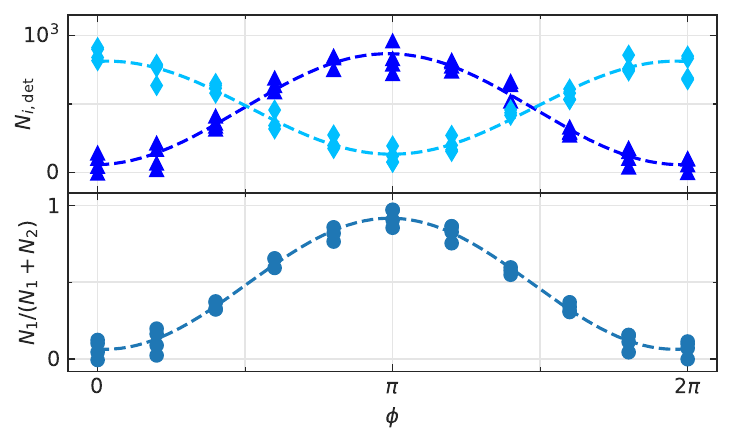}
	\caption{Representative coherent oscillations induced by a spin-echo sequence to determine the imaging calibration factors. The upper panel shows the raw atom numbers extracted from absorption images for Cs$\ket{1}$ (dark blue triangles) and Cs$\ket{2}$ (light blue diamonds)%
		. The lower graph is calculated from $N_1$ and the corrected $N_2$ as described in the main text. The dashed lines are sinusoidal fits to the respective data.
	}
	\label{fig:CsDoubleImg_oscillations}
\end{figure}

Using this sample, consisting of a majority population in Cs$\ket{2}$ and a small parasitic population in Cs$\ket{1}$, we perform spin-echo measurements \cite{Hahn1950_Spin} to induce coherent oscillations between the two states. This is done by a first $\pi/2$-pulse which brings the 
Cs$\ket{1}$ and the Cs$\ket{2}$ states into an equal superposition. After half of the free evolution period a $\pi$-pulse is introduced, and at the end of the free evolution another $\pi/2$-pulse with a variable phase $\phi$ is applied. This protocol is intrinsically robust against our dominant source of technical noise - magnetic-field fluctuations - and is therefore particularly well suited for this purpose. The pulses for the spin-echo protocol are performed using the Raman setup with a Rabi frequency of $\Omega_\mathrm{R} = 2 \pi \times \SI{500}{\kilo \hertz}$, %
while the chosen method to perform the Cs$\ket{1}$\;$\leftrightarrow$\;Cs$\ket{2}$ transfer for the imaging sequence is a $\pi$ pulse with the same Rabi frequency. An example of the observed coherent oscillations as a function of the phase $\phi$ of the second $\pi/2$ pulse is shown in the upper panel of Fig.~\ref{fig:CsDoubleImg_oscillations}.

To determine the imaging calibration factors, we use five datasets recorded on different days. Each dataset typically comprises 35 measurements, with each measurement consisting of approximately 44 data points sampling one sinusoidal oscillation at a fixed free evolution period. Sampling a single oscillation requires, in this case, approximately \num{20}\,min, which averages over fluctuations between individual data points. Since a complete dataset is acquired over approximately \num{12}\,h, it also captures  slow drifts. The calibration is performed assuming the heuristic relation
\begin{equation}
	N_{2,\,\mathrm{det}} = c_1 N_{2} + c_2 \quad \mathrm{,}
\end{equation}
which connects the atom number in the Cs$\ket{2}$ state to the atom number $N_{2,\,\mathrm{det}}$ that can actually be detected. While the efficiency parameter $c_1$ is assumed to be constant across all measurements, the offset parameter $c_2 = \tilde{c}_2 \times \overline{N}$ is expected to scale with the total number of Cs atoms $\overline{N}$. Since the measured total atom number is also subject to noise, we use its average $\overline{N}$ over all data points comprising one oscillation, acquired within a recording time of approximately \num{20}\,min.

Each of the recoded oscillations is fitted individually with a sinosodial function. The efficiency parameter $c_1$ is then given as the ratio of the amplitudes from the oscillations in $N_{2,\,\mathrm{det}}$ and $N_{1}$. This follows directly from the assumption that only two states are involved and the observation that losses are negligible during the probing interval, ensuring conservation of the total Cs atom number. After averaging over all measurements and datasets, we end up with $c_1 = \num{0.85(3)}$. %
Due to the dependence of the offset parameter $c_2$ on the absolute atom number, it needs to be quantified recursively. We determine this parameter by requiring each individual oscillation experiment to yield a sinusoidal curve centered at $N_1/(N_1 + N_2) = 0.5$, which follows from conservation of the total Cs atom number. Using this protocol, we determine the reduced offset parameter $\tilde{c}_2$. Averaging over all individual oscillations yields $\tilde{c}_2 = \num{0.059(0.012)}$. %
Interestingly, we find no evidence of correlations between the imaging parameters $c_1$ and $\tilde{c}_2$ and either the measurement time or the total Cs atom number. In addition, no statistically significant differences are observed between datasets acquired on different days, indicating that the uncertainties of the imaging parameters are dominated by shot-to-shot fluctuations.

It is important to note that the efficiency parameter $c_1$ is determined predominantly by losses during the imaging process rather than by the finite transfer efficiency between Cs$\ket{1}$ and Cs$\ket{2}$. As discussed in the main text, optical pumping into the $\ket{F=4,m_F=4}$ state is required to drive the closed $\ket{F=4,m_F=4} \leftrightarrow \ket{m_I^\prime=7/2,m_J^\prime=3/2}$ imaging transition. Despite a detuning of approximately \SI{230}{\mega \hertz} from the Cs$\ket{2} \rightarrow \ket{F=4,m_F=4}$ transition, the finite linewidth of the home-built distributed Bragg reflector laser, approximately \SI{60}{\mega \hertz}, results in off-resonant photon scattering losses of about \SI{10}{\percent} in the Cs$\ket{2}$ population. Despite extensive investigations, we were unable to identify the origin of the finite value of $c_2$. We can exclude the Cs$\ket{1}$\;$\leftrightarrow$\;Cs$\ket{2}$ transfer process and the optical-pumping process. Additionally, neither the light driving the closed imaging transition nor camera-related artifacts appear to be responsible.

Having characterized the imaging process, we can now assess the precision with which the transferred fraction is determined. The five complete data sets are analyzed and converted into transferred fractions using the previously determined imaging parameters. Each oscillation is fitted individually with a sinusoidal function, and the deviations of the measured data points from the corresponding fit are used to estimate the measurement uncertainty. The resulting distribution is shown as the blue histogram in Fig.~\ref{fig:CsDoubleImg} of the main text. For comparison, we define a pseudo-transferred fraction as $N_1/N_{1,\mathrm{max}}$, where $N_{1,\mathrm{max}}$ is obtained from the maximum of the fitted sinusoid (gray in Fig.~\ref{fig:CsDoubleImg}). As mentioned in the main text, we find a significant reduction in the determination uncertainty of the transferred fraction when both Cs states are taken into account.

\FloatBarrier
\bibliography{main.bib}%

\begin{thebibliography}{93}
\bibitem [{TAt()}]{TAtoptica}%
  \BibitemOpen
  \href@noop {} {\bibinfo {title} {{Data replotted with kind permission from:
  Production and Quality Control Data Sheet, TOPTICA TA pro 850, measured with
  an optical spectrum analyzer (OSA) with a resolution of \SI{0.05}{\nano
  \meter}. The amplified diode laser used for the remaining measurements has a
  background suppression of $\SI{43}{\dB}$.}}}\BibitemShut {Stop}%
\end{thebibliography}

\end{document}